\documentclass[10pt,letterpaper]{article}
\usepackage{opex3}

\begin{document}

\title{A new bound on excess frequency noise in second harmonic generation in PPKTP at the $10^{-19}$ level}
\author{D. Yeaton-Massey$^{1,*}$ and R. X. Adhikari$^1$}
\address{$^1$California Institute of Technology, Division of Physics, Math, and Astronomy \\ Pasadena, CA, 91125, USA}
\email{$^*$davidym@caltech.edu}

\begin{abstract}
We report a bound on the relative frequency fluctuations in nonlinear second harmonic 
generation. A 1064$\,$nm Nd:YAG laser is used to read out the phase of a Mach-Zehnder 
interferometer while PPKTP, a nonlinear crystal, is placed in 
each arm to generate second harmonic light. By comparing the arm length difference of the 
Mach Zehnder  as read out by the fundamental 1064~nm light, and its second 
harmonic at 532~nm, we can bound the excess frequency noise 
introduced in the harmonic generation process. We report an amplitude spectral density 
of frequency noise with total RMS frequency deviation of $3\,$mHz and a minimum 
value of 20 $\mu{\rm Hz}/{\rm Hz}^{1/2}$ over 250 seconds with a measurement bandwidth of 
128~Hz, corresponding to an Allan deviation of $10^{-19}$ at 20 seconds.  
\end{abstract}

\ocis{(190.2620) Harmonic generation and mixing; (120.2920) Homodyning; (120.3180) Interferometry. }

\bibliographystyle{osajnl}

\section{Introduction}
Harmonic generation of optical fields in nonlinear optical crystals can be modeled with classical 
coupled wave equations~\cite{Bloem:1962}. Define the fundamental and second harmonic fields as 
$E_1=\mathcal E_1 \exp(2 \pi i \nu_1 t)$ and $E_2=\mathcal E_2 \exp(2 \pi i \nu_2 t)$, respectively. In 
ideal second harmonic generation,  $\nu_2 = 2\nu_1$. While this frequency ratio has been measured very 
precisely ($\langle\nu_2/\nu_1\rangle=7\times10^{-19}$ \cite{Stenger:2002}), we present a lower bound to 
the frequency noise, or the spectrum of the time dependent quantity $\nu_2 - 2\nu_1$.

Several experiments at the forefront of precision metrology and frequency standards use harmonic 
generation in their experiments. These include iodine stabilized Nd:YAG lasers~\cite{Zang:2007}, optical 
frequency combs~\cite{Grosche:2008,Coddington:2007}, measurement of optical frequency 
ratios~\cite{Stenger:2002,Rosenband:2008}, and precision atomic spectroscopy~\cite{Fortier:2007}.  Many of 
these experiments provide bounds to any excess frequency noise which might be found in the second harmonic 
generation (SHG) process (e.g. from thermodynamic fluctuations in the crystal temperature). As experiments like these 
push towards lower noise levels, the fundamental noise sources in second harmonic generation may become a 
relevant noise source.

Another field which utilizes precision SHG is interferometric detection of gravitational 
waves. These interferometers require 
measurements of mirror displacements at the level of 10$^{-20}\,{\rm m}/\sqrt{\rm Hz}$~\cite{LSC:2009}. 
To achieve such high precision, these interferometers use seismically isolated suspended mirrors which
make up multiple coupled optical cavities with narrow linewidths. It has proven challenging to control these
cavities and bring them simultaneously into resonance. In order to simplify this problem, auxiliary fields
will be generated by SHG and used to form independent signals for two of the cavities.
This technique is being developed using some small scale prototypes~\cite{Mullavey:12, Izumi:2012}. 
To understand the limits of such a scheme, we need to bound the frequency noise introduced in the second 
harmonic generation process. In our case, we use a 1064~nm carrier (Nd:YAG), and generate the 532~nm 
doubled light with PPKTP crystals. In order to meet the stability requirements for this technique,
the excess frequency noise must be less than 70~mHz RMS from 10~mHz 
to 30~Hz~\cite{Mullavey:12}. In the future, optical configurations using more than one wavelength inside 
the interferometer to beat standard quantum noise limits 
of the detection~\cite{Khalili:2011} may be used. Fundamental noise limits in harmonic generation may be 
relevant there.

\section{Experimental Setup}

\begin{figure}[h]
\centerline{
    \includegraphics[width=1\columnwidth]{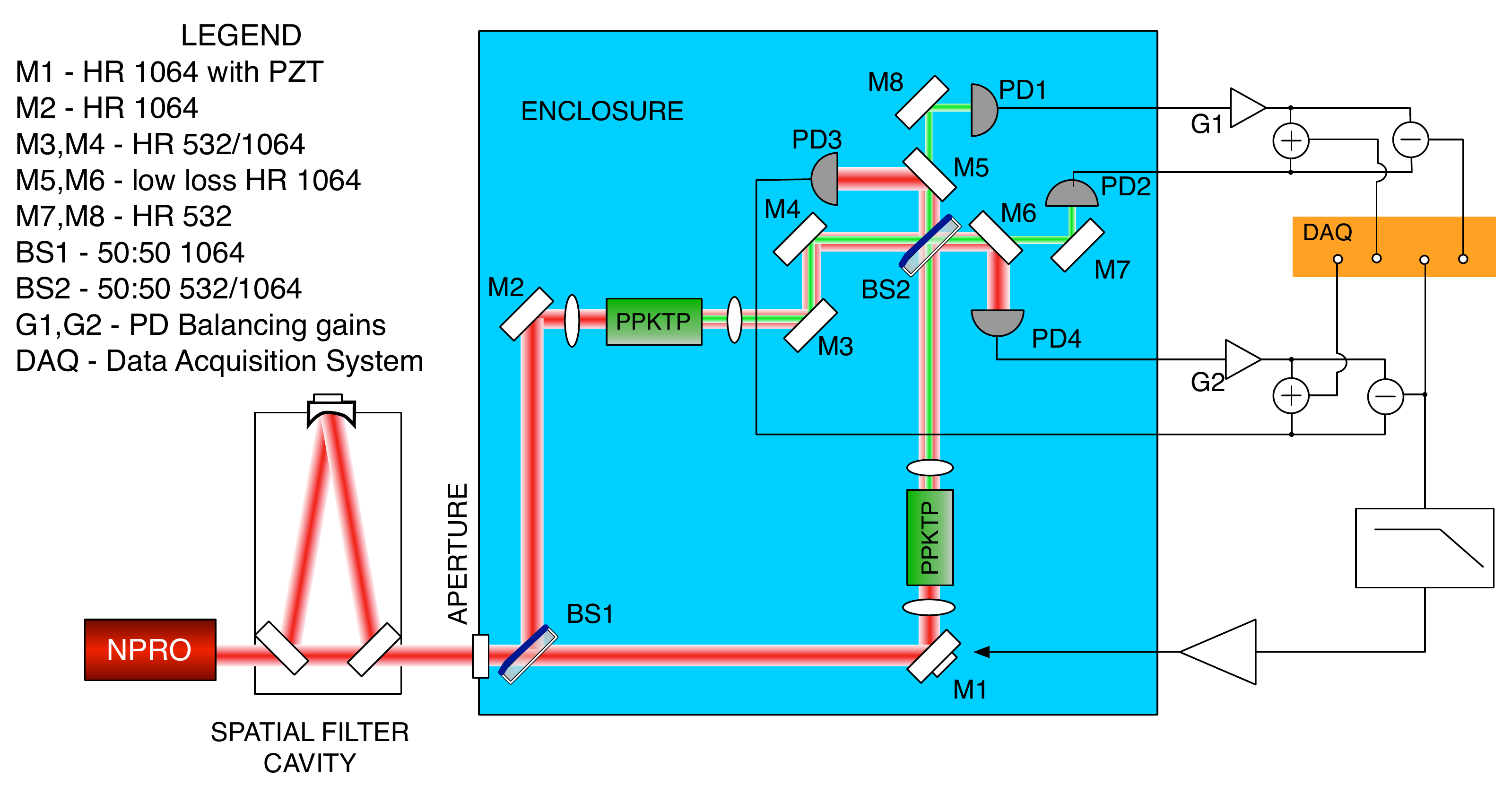}}
    \caption{Experimental layout - A dual wavelength Mach-Zehnder Interferometer. Some readout mirrors for 
               harmonic separation are not shown. See text for a more complete description.}
\label{fig:diagram}
\end{figure}

Figure~\ref{fig:diagram} shows the experimental setup used to measure the uncorrelated frequency noise between 
the fundamental and the second harmonic. The beam from a 2\,W non-planar ring laser at 1064~nm is passed 
through a spatial filter cavity~\cite{Benno:PMC}. The spatially filtered beam then enters the Mach-Zehnder through a 
hole in an acoustic enclosure. PPKTP crystals (Raicol Crystals) inside temperature stabilized ovens
are placed in each arm, with lenses added to mode match to the each crystal. Two additional 
dichroic mirrors (M3, M4) are placed in one arm for alignment. The beams recombine on a dichroic 50:50 beamsplitter 
and low loss HR mirrors (T $<$ 10~ppm at 1064~nm) are used to separate the fundamental from 
the second harmonic. Commercial dichroic mirrors (HR532, AR1064) were used to further separate the 532~nm from the 
1064~nm light. For the 532~nm and 1064~nm detection, 
Si (Hamamatsu 1223) and InGaAs (GPD 2000) photodiodes were used, respectively.
The Mach-Zehnder arm lengths were adjusted to be mid fringe for both 532 and 1064~nm simultaneously. 
The interferometer is locked to this point by applying feedback from the 1064~nm PD difference 
signal to the PZT on M1 (see Fig.~\ref{fig:diagram}). We drove the PZT with a small signal at 100~Hz to monitor the 
calibration (rad/V) at both wavelengths, as any differential frequency noise between the two wavelengths would cause the 
532~nm calibration to drift. The calibration of the Mach Zehnder was confirmed by sweeping 
through multiple fringes. The 1064~nm and 532~nm Mach-Zehnder length signals were read out 
with two pairs of balanced homodyne detectors. The sum and difference signals were digitized at 
256~Hz and processed further offline.

In the non-depleted pump approximation, with imperfect phase 
matching, the phase relation of the two fields in each arm before the 
recombining beamsplitter is:

\begin{equation}
\theta_{532} = 2\theta_{1064}-\pi/2-\Delta k L/2\,,
\end{equation}
where $\theta_{1064}$ and $\theta_{532}$ describe the relative phases of the fundamental and the second harmonic, 
respectively, $L$ is the length of the doubling crystal, and $\Delta k$ is the phase mismatch, the parameter normally 
used to describe the efficiency of SHG. 
In theory, $\Delta k$ can be arbitrarily small (limited in practice only by the ability to stabilize 
the temperature of the nonlinear crystal). 

Using superscripts to differentiate between the two arms, the difference in the second harmonic phase is thus
\begin{equation}
\theta_{532}^{A}-\theta_{532}^{B} = 2\left(\theta_{1064}^{A}-\theta_{1064}^{B}\right)-\left(\Delta k^{A}-\Delta k^{B}\right)L/2\,.
\end{equation}

\section{Results}
\label{sec:Res}

\begin{figure}[h]
\centerline{
\includegraphics[width=0.7\columnwidth]{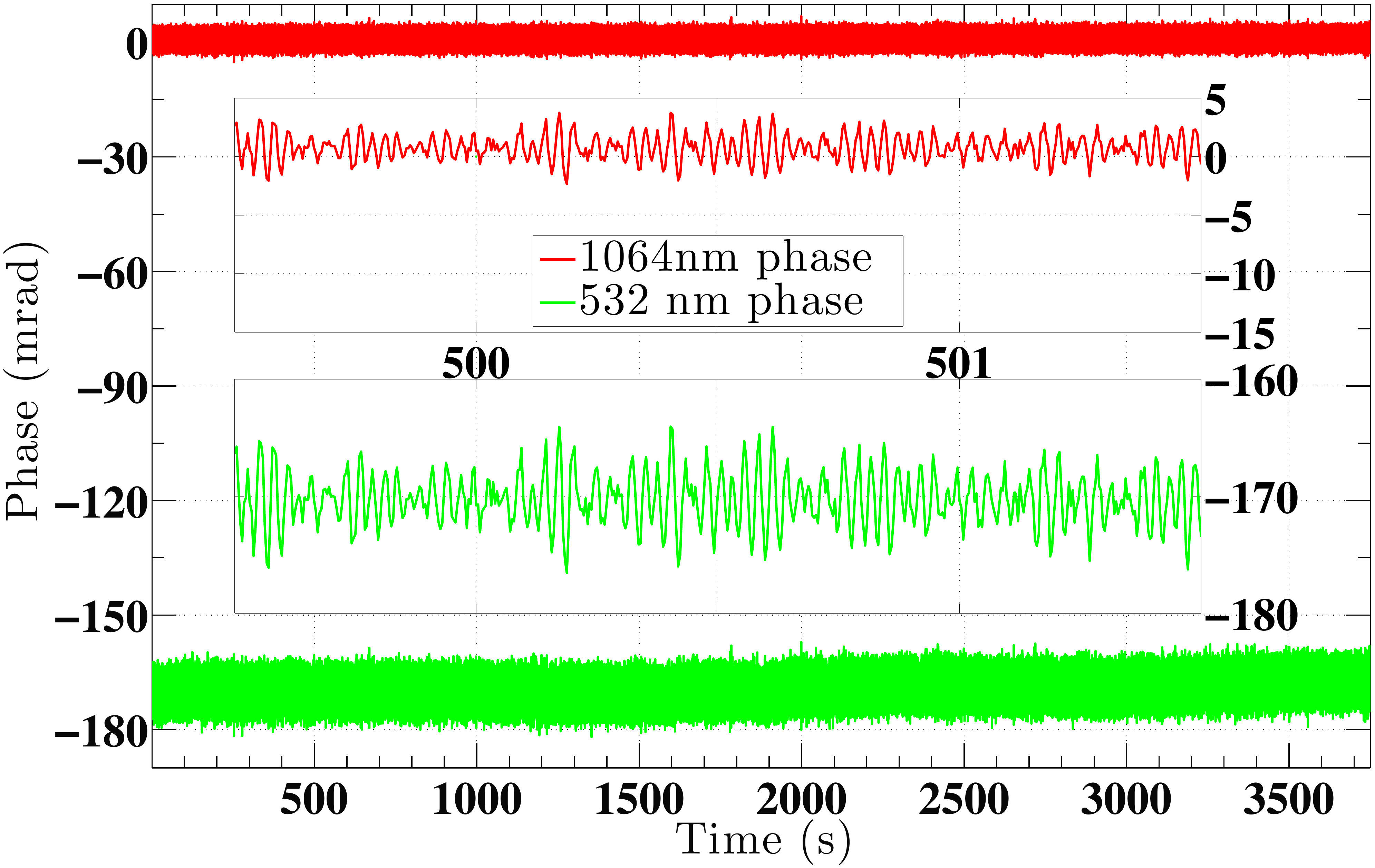}}
\caption{Example time series of the Mach-Zehnder output. The RMS phase noise over the one hour period was 
6~mrad RMS at 532~nm and 3~mrad RMS at 1064~nm}
\label{fig:PHASE}
\end{figure}

The phase difference ($ \delta \theta_{1064} (t)\equiv \theta_{1064}^{A}(t)-\theta_{1064}^{B}(t)$) was suppressed by 
the servo, which had a unity gain frequency of $\sim$~10~Hz.
Typical values of $\delta\theta_{1064}(t)$ and $\delta\theta_{532}(t)$ are shown in Fig.~\ref{fig:PHASE}.

\begin{figure}[h]
\centering
	\includegraphics[width=\columnwidth]{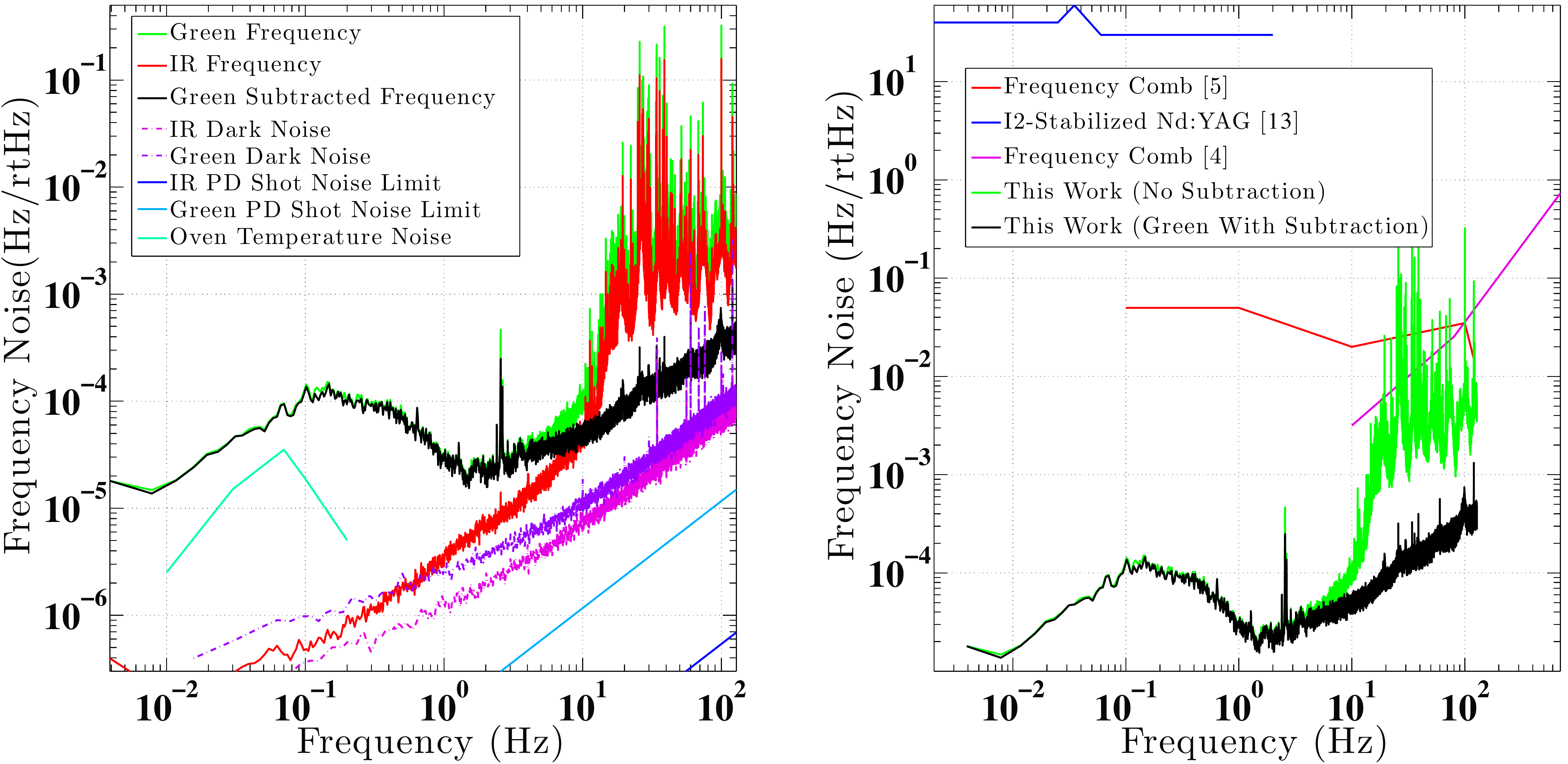}
     \caption{On the left we see the full noise budget of the experiment:  $\theta_{1064}^{A}-\theta_{1064}^{B}$ (in red) 
                  (which is the in-loop signal of the Mach-Zehnder), $\theta_{532}^{A}-\theta_{532}^{B}$ (in green) (which contains 
                   excess phase noise), and the subtraction residual (in black). The total RMS frequency 
                  fluctuation in the measurement band is 3~mHz. On the right we have a comparison of 
                this work with previous bounds. These bounds show up as frequency 
              noise~\cite{Leonhardt:2006}, timing error \cite{Coddington:2007}, and phase
              noise~\cite{Grosche:2008}.}
\label{fig:NB}
\end{figure}

The frequency noise amplitude spectral density of the Mach Zehnder is shown in Fig.~\ref{fig:NB} with known 
noise sources. $\delta \nu_{1064}$ and 
$\delta \nu_{532}$ are shown in red and green, respectively, where $\delta \nu \equiv \delta \dot{\theta}/2\pi$. 
Above 10~Hz the phase noise is dominated by a forest of mechanical resonances on 
the optical table which show up strongly in both $\delta \nu_{1064}$ and $\delta \nu_{532}$. 
We used the average transfer function over the measurement time, 
$\textrm{H}(f)=\langle\delta \nu_{532}(f)/\delta \nu_{1064}(f)\rangle$, to subtract
the noise which is coherent between the 
two $\delta \nu$s. This subtracted level is shown as the black trace in Fig.~\ref{fig:NB}. 
The black trace bounds any noise source which causes frequency noise between the 
fundamental and the harmonic 
and is uncorrelated between the two ovens, such as thermodynamic fluctuations in the crystals, or any 
temperature fluctuation not common to both ovens. Some common mode effects such as intensity 
dependent phase shifts and temperature 
fluctuations (and thus phase matching fluctuations) are suppressed by the experimental setup, so
these technical noise sources will not be visible in these measurements. However, a pessimistic estimate 
of the temperature noise coupling assuming no common mode rejection shown in Fig.~\ref{fig:NB} is below 
the measured excess frequency noise. In addition, the intensity of the fundamental was only stabilized to a 
level of $2\times 10^{-6}$ at 1 Hz, where it can be lowered to the $10^{-8}$ at 1~Hz with current techniques, so 
unless the rejection of this effect was more than 46 dB, we can safely ignore it. It is highly improbable that 
the intensity to frequency noise coupling would be above the level shown in Fig.~\ref{fig:NB}.
 The excess noise 
below 10~Hz was found to be correlated with air currents on the table, and would be reduced by 
moving the setup into a vacuum chamber.
The total RMS excess frequency noise of the black trace is 3~mHz RMS in the 
10~mHz to 128~Hz band.

\section{Noise Sources}
\label{sec:noise}
In addition to the usual technical noise sources, it is worth considering whether there is a more fundamental
limit to the relative phase between the fundamental and the harmonic. 
A rough estimate of thermal noise from thermoelastic (Zener) damping was obtained by 
directly applying~\cite{Levin:2007} the Fluctuation-Dissipation Theorem. We treat the crystal as 
an 0.8~mm radius cylinder, and follow the calculation done by Heinert et al.~\cite{Heinert:2011}. 
This yields a spectral density of $\Delta k$ taking into account both thermorefractive and thermoelastic 
fluctuations. When expressed as frequency fluctuations, it is well approximated by 
$2.5 / (1+500 f^{-7/8})\,\mu\textrm{Hz}/\sqrt{\rm Hz}$ above 10~Hz. 
Below that, it must be flat or continue to decrease, or else the RMS temperature integral would diverge. 
Practically speaking, in our band of interest, the temperature fluctuations of the ovens is at least 2 orders 
of magnitude greater than these fundamental thermodynamic temperature fluctuation. In the future,
when researchers seek to make frequency comparisons at better than the $10^{-21}$ level, these
thermal noises will have to be calculated with more accuracy.

\section{Previous Bounds on Excess Frequency Noise}
\begin{figure}[h]
\centerline{
\includegraphics[width=0.8\columnwidth]{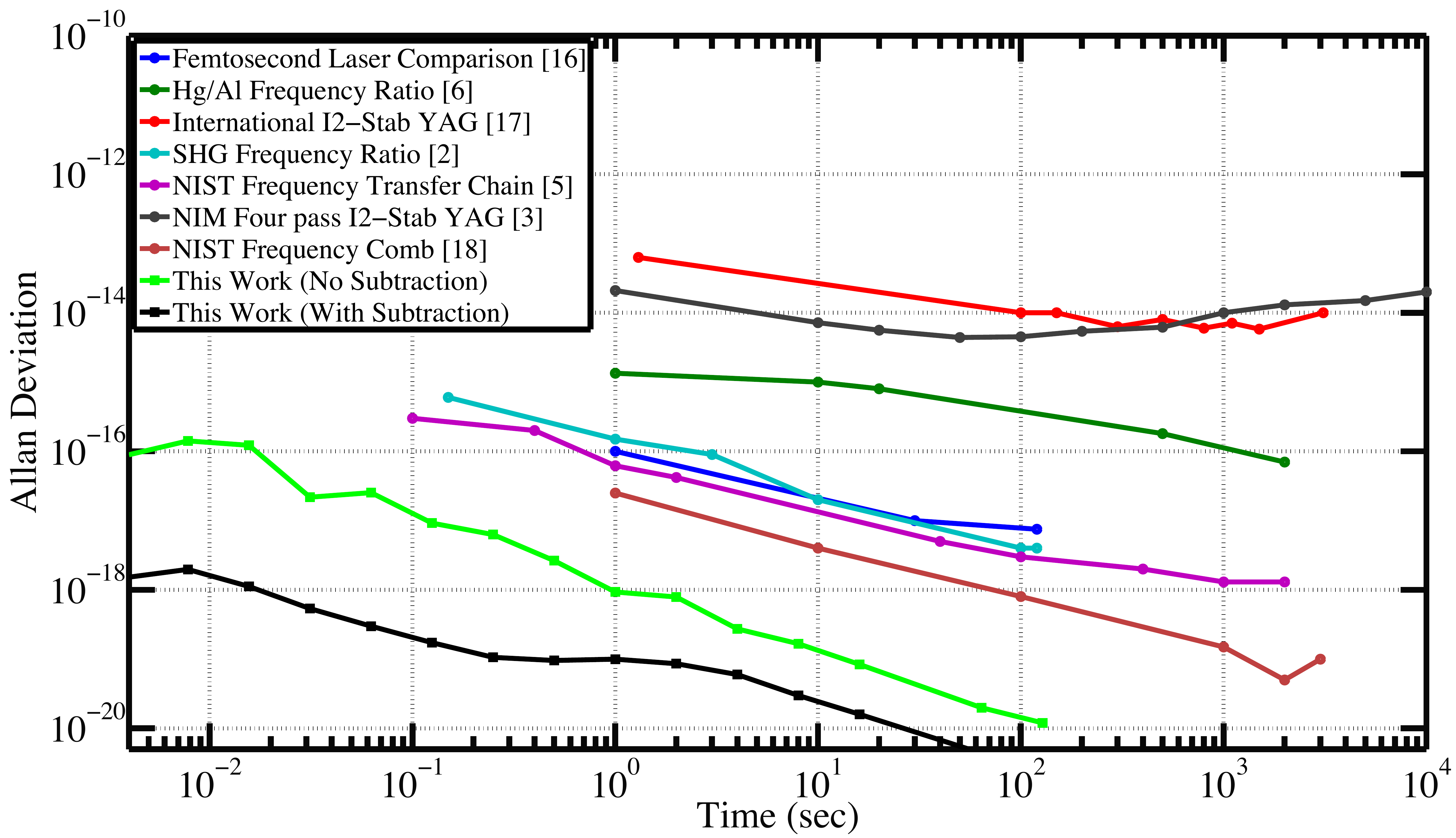}}
\caption{The Allan deviations from this experiment (green and black) were obtained from 
               the power spectrum as described in \cite{CCIR:1986}. Bounds from previous 
               work are shown for reference.}
\label{fig:AVAR}
\end{figure}

We have examined the results from a number of precision experiments which involve SHG.
Although these experiments are likely limited by other noise sources, these data can be used to 
place upper bounds on the possible noise generated by SHG.
In Fig.~\ref{fig:NB} we compare frequency noise~\cite{Leonhardt:2006}, timing 
errors~\cite{Coddington:2007}, and phase noise~\cite{Grosche:2008}; in Fig.~\ref{fig:AVAR} we 
compare Allan 
deviations~\cite{Stenger:2002,Zang:2007,Coddington:2007,Rosenband:2008,LSMa:2004,Robertsson:2001,Stalnaker:2007}. 
While the comparison in Fig.~\ref{fig:NB} is straightforward, some caution should 
be taken interpreting Fig.~\ref{fig:AVAR}. Our Allan deviation at the 0.1~s time scale is 
heavily influenced by the high frequency noise in the measurement (10-128~Hz). Since 
we low pass the signal to acquire data at 256~Hz, we reject noise which would make the 
Allan deviation increase at all time scales. It should also be emphasized that we only 
measure relative frequency fluctuations, and that there are some common mode noise 
sources which the experiment is insensitive to. See~\cite{Rubiola:2008} for more information on 
Allan deviations and phase noise.

\section{Conclusions} 
In conclusion, we have demonstrated that the RMS frequency fluctuations added from uncorrelated 
mechanisms between two SHG crystals is less than 3$\,$mHz at time scales over 10~ms. The obvious 
correlated mechanisms (temperature and intensity noise coupling) are likely insignificant compared to 
this as discussed in section~\ref{sec:Res}.
This is low enough to not limit the lock acquisition~\cite{Mullavey:12, Izumi:2012} 
scheme for Advanced LIGO and other gravitational-wave detectors. Additionally, we have shown that 
there is no excess noise process
at a level which is of interest to those doing precision atomic spectroscopy~\cite{Fortier:2007}, 
using frequency combs to transfer optical harmonics~\cite{Grosche:2008,Coddington:2007,Terra:2009} 
and other tests of fundamental physics.

\section*{Acknowledgements}
The authors would like to thank P. Fritschel for suggesting this work as well as K. Arai, J. Harms, 
and K. McKenzie, for discussions leading to the understanding presented in this letter, as well as 
H. Grote and M. Evans for useful input on the interferometer stabilization.
We also gratefully acknowledge the National Science Foundation for support under 
grant PHY-0757058.

\end{document}